Opinion Article
We Did See This Coming:
Response to, "We Should Have Seen This Coming," by D. Sam Schwarzkopf

**Authors and affiliations**
Julia Mossbridge, M.A., Ph.D., Northwestern University
  (corresponding author: j-mossbridge@northwestern.edu)
Patrizio Tressoldi, Ph.D., Università di Padova
Jessica Utts, Ph.D., University of California at Irvine
John A Ives, Ph.D., Samueli Institute
Dean Radin, Ph.D., Institute of Noetic Sciences
Wayne B Jonas, M.D., Samueli Institute

We appreciate the effort by Schwarzkopf to examine alternative explanations for predictive anticipatory activity (PAA) or presentiment (for first response, see: Schwarzkopf 2014a; for additional response, see: Schwarzkopf 2014b, for original article, see: Mossbridge et al. 2014). These commentaries are a laudable effort to promote collegial discussion of the controversial claim of presentiment, whereby physiological measures preceding unpredictable emotional events differ from physiological measures preceding calm or neutral events (Mossbridge et al., 2012; Mossbridge et al., 2014). What is called truth at any given time in science has achieved that status through a continuous process of measurement and interpretation based on the current knowledge at hand. Here we address six points in Schwarzkopf's original commentary (Schwarzkopf 2014a), though our responses are informed by the comments in his supplementary commentary (Schwarzkopf 2014b). We hope our responses will help Schwarzkopf and others understand our interpretation of these data.

Point 1: "Any meta-analysis can only be as good as the studies it analyzes…several of the studies included are of questionable quality, e.g. the fMRI experiment (Bierman and Scholte, 2002)…moreover, many studies were published in conference proceedings and did not pass formal peer review…additional factors would have been of importance, such as whether the experimenters expected to find a presentiment effect."

For these and other reasons, the results of removing the Bierman and Scholte study (2002) as well as any other study that did not use skin conductance as a physiological measure were already reported in the original meta-analysis (Mossbridge et al. 2012). The results were still overwhelmingly statistically significant. As to the point regarding conference proceedings, all of the conference proceedings included in that meta-analysis passed formal peer review (Mossbridge et al. 2012). As a cultural note, those of us who have submitted papers to mainstream and to parapsychology conferences know that the peer review process for parapsychology is generally more rigorous than for other conferences. One paper was not submitted to a conference but was instead a research report (Bierman 2007); the effect size from that study went against the hypothesis of the meta-analysis. Finally, in regards to whether the experimenters expected to find an effect, it is not clear how such an expectation would have influenced the results, and in any case researchers always have some expectation in mind when they conduct experiments.

Point 2: "The meta-analysis should be much broader, including myriad studies not conducted by psi researchers…An often-used argument, that these studies are invalid because they used counterbalanced designs and are thus confounded by expectation bias, is a straw man and also rather ironic given the topic of investigation: unless participants knew in advance that stimuli were counterbalanced this could not possibly change their expectations."

The first sentence is a non sequitur. Anyone who studies presentiment may be described as a psi researcher, i.e. someone who studies a purported psi effect. So broadening the meta-analysis to non-psi-researchers doesn't make sense.

When participants make guesses at future events in binary sequences that are believed to be random, they tend to discount the possibility of long outcome runs (> 3) in a random sequence (*e.g.,* Ayton and Fischer 2004), demonstrating a clear expectation that after a few similar outcomes, the outcome will change (*e.g.,* the "gambler's fallacy"). This expectation grows with the length of the outcome run (Boynton 2003). This is the case regardless of whether stimuli are sampled from a dataset with or without replacement.

However, when stimuli are sampled without replacement, the probability of stimulus type A being chosen on trial n+1 depends on the proportion of each type chosen on previous trials. Statistically, a stimulus that has not shown up so much in the past sequence is more likely to show up in the future sequence if the stimuli are sampled without replacement, as is normally done in most mainstream psychophysiology and neuroscience laboratories. In this case, the concern is that when stimuli are sampled without replacement, the gambler's fallacy is more likely to lead to a false presentiment effect. In contrast, most presentiment researchers sample stimuli with replacement. Further, they use hardware-based true random number generators (RNGs) as opposed to pseudorandom algorithms, ensuring that there is no way for participants or investigators to predict the upcoming stimulus type. Although it is highly unlikely that the sequences produced by pseudo-RNGs can include sequences that have recognizable patterns, some of the authors (JM, JU, DR and PT) have had manuscripts based on presentiment critiqued and/or rejected for use of a pseudo-RNG. Further, in the past, reviewers have attributed presentiment effects in mainstream research to participants learning a subtle pattern in the sequence. Thus we must err on the side of caution.

Even so, in the original meta-analysis we reported analyses of the only data we could obtain from mainstream psychophysiology labs (two out of fourteen data requests). In fact, we did find effects similar to presentiment, but we did not conclude much due to the reasons explained above (Mossbridge et al. 2012).

Point 3: "A particular critical factor that should have been analyzed directly is the imbalance between control (calm) and target (arousing) trials typically used in these studies…this is usually approximately 2:1…such an imbalance means that an attentive participant will quickly learn statistical properties of the sequence and thus affects how the brain responds to the different stimulus classes."

To reject the null hypothesis that temporally "downstream" events (physiological differences after emotional versus neutral stimuli) are not anticipated by "upstream" events (physiological differences before emotional versus neutral stimuli), it is critical to have a clear physiological difference following emotional versus neutral stimuli. Post-stimulus responses to emotional stimuli habituate with continued target presentation (*e.g.,* Bradley et al. 1993), so a greater neutral-to-emotional stimulus ratio is common among PAA experiments. However, the chance of the next event being a target in a series sampled with replacement is 50% with a 1:1 ratio of non-target to target stimuli, 33.33% with a 2:1 ratio, and 25% with a 3:1 ratio. Although this imbalance will bias expectations, it will bias them in the wrong direction to produce an effect. Specifically, if the participant realizes ¾ of the trials are target trials, then the expectation on each trial should be for neutral trials (aside from the gambler's fallacy, discussed above). This will make finding a physiological difference between neutral and target trials *less* likely. In other words, a brain that has learned that an emotional stimulus is not as likely as a calm one would do best to predict a calm stimulus on each trial, making it *more* impressive that there is a significant *difference* between the activity preceding calm and emotional trials in experiments using imbalanced stimulus ratios.

Point 4: "Could these effects be at least partially explained by analytical artifacts? In many of these studies (Bierman and Scholte, 2002; Radin, 2004) the data are not only baseline corrected to the mean activity level prior to stimulus onset, but they are further "clamped" to a particular time point prior to the stimulus."

Having addressed the Bierman and Scholte (2002) study (point 1), here we address the Radin (2004) study. The method of baseline correction is *z*-transformation based on the mean and standard deviation of the values in the pre-stimulus period. By definition, this produces a difference of zero in the pre-stimulus period between all traces used in the *z*-transformation. To examine differences in pre-stimulus activity, it is thus necessary to clamp the data. The purpose of baseline correction is to reduce the possibility of influence on the current trial from lingering effects of the previous trial. Because the clamping is done *after* normalization, it is not clear how the previous stimulus could be influencing the data. However, with other baselining methods the previous stimulus could influence the baseline, a problem that would be more of a concern if stimuli were not randomized, were sampled without replacement, if there was insufficient time for physiological responses to return to baseline between stimuli, or if stimulus sequences contained only a few trials. In addition, Radin's statistical analyses used randomized permutation methods, which make no parametric assumptions and account for autocorrelations in the data (Blair and Karniski 1993).

Point 5: "Many of these studies test for the presence of expectation bias by correlating the presentiment effect with the time between target events (Mossbridge et al. 2012)…this is based on an unproven assumption that these physiological effects scale linearly with expectation. Further, because the probability of sequences of control trials falls off exponentially with their length, the presentiment effect cannot be estimated with the same reliability for long sequences as for short ones…Crucially, does presentiment persist when participants do not expect a target even when the next trial is one?"

We agree that testing for expectation bias and effects of trial order is critical (Dalkvist and Westerlund 2006; Kennedy 2013; Dalkvist et al. 2014). To address the problem of longer sequences, some of us have disregarded longer lags (Mossbridge et al. 2011), although in Radin (2004) even a cursory look at only the first five lags indicates that the presentiment effect is not due to expectation bias.

In our simulations of linearly-produced expectation bias (details in Matlab code posted at https://www.dropbox.com/s/6gfnok34felee2m/SimulatingExpBiasPAA.m), expectation bias analyses seem to catch most of the expectation-induced presentiment effects as the number of participants increases (Figure 1). An average of 35 participants were used in the 19 studies using expectation bias analyses (Mossbridge et al., 2012). Based on this simulation, around 8% of these studies could have undetected false presentiment effects induced by expectation bias, leaving around 92% of these results unexplained. Further, if expectation bias explained the presentiment effect among these 19 studies, there should be a significant negative correlation between effect size and number of participants, which was not the case ($r = -0.16$, $p > 0.524$).

Schwarzkopf correctly points out that expectation bias may be non-linear. Indeed, most physiological signals do not scale linearly with expectation because at some point arousal "tops out." In our opinion, the best approach to handling order effects is the one we outlined in the original meta-analysis: perform single-trial studies with between-participant comparisons to look for presentiment effects that, if significant, cannot be due to trial order or traditional expectation bias (see Figure 6 in Mossbridge et al. 2012). Schwarzkopf proposes a false-cuing experiment, but this suggestion is based on the assumption that if the presentiment effect exists, it would trump the cuing effect. This is unlikely given the generally small effect size.

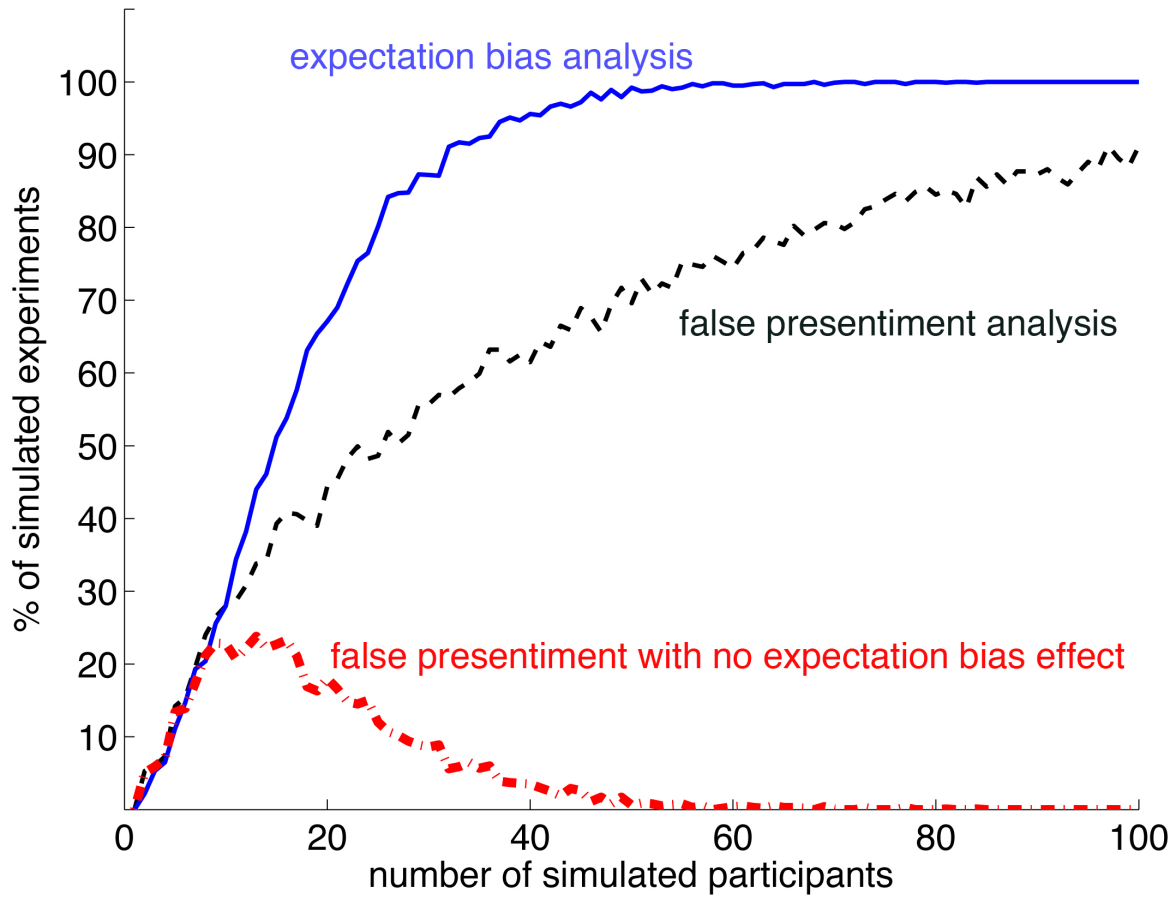

**Figure 1.** Results from 1000 stimulated experiments with each of 100 different numbers of simulated participants. The y-axis gives the percentage of simulated experiments that produced results according to three measures: 1) expectation bias analysis (blue solid line), which for each experiment determines whether there is a significant linear correlation ($p<0.05$) between the number of calm trials preceding an emotional trial (1-5 preceding) and the pre-stimulus activity, 2) the false presentiment analysis (black dashed line), which for each experiment determines whether there is a significant paired t-test ($p<0.05$) between pre-stimulus activity preceding emotional versus calm trials (this is the standard comparison to establish true presentiment effects, but here the only such effect is induced by a simulated expectation bias), and 3) cases in which there is a significant false presentiment effect and no detected expectation bias effect (red dotted line). Note that cases in which the expectation bias analysis does not catch the false presentiment effect dwindle with increasing numbers of participants. Matlab code available at: https://www.dropbox.com/s/6gfnok34felee2m/SimulatingExpBiasPAA.m.

Point 6: "Lastly, are the purported effects even biologically plausible?...If these responses themselves reverse the arrow of time and are caused by future stimuli, this will require a complete overhaul of current theory."

We do not propose that these responses reverse the arrow of time, but rather that they may reflect time-symmetric processes that are known to be an inherent part of physical reality at the microscopic scale. While the effect measured via psychophysiological signals appears at first glance to be macroscopic, the underlying cause of those physiological effects may well be microscopic, e.g., a few synapses in the brain. As to the point that these results would invalidate other scientific results, this may be the case for some theories. But theories regularly evolve and expand to accommodate new data, so this should not be surprising nor threatening. As we stated in the original meta-analysis and in the follow-up review, we are not proposing a theory about the underlying mechanisms. However, the meta-analytic results were highly significant and homogenous (Mossbridge et al. 2012, Mossbridge et al. 2014). We looked at factors that could influence the quality of evidence in the original meta-analysis, and we found that higher-quality studies produced a higher effect size (Mossbridge et al. 2012).

We would like to point out that time reversal might seem like a strange concept, but it is not inconsistent with well-established physics. As in our review, we again point to an excellent discussion available at: http://plato.stanford.edu/entries/causation-backwards/.

It is a common misconception to assume that what we already know about conscious temporal flow also applies to the temporal flow of information encoded by non-conscious processes. To actually test this assumption, one must look at physiological measures both before and after stimuli, as is done in presentiment experiments, or in re-analyses of data recorded for other purposes. When such analyses are performed, as we have shown, there is a clear result: non-conscious physiological changes occur both before and after stimuli are presented (Mossbridge et al. 2012). We are aware of no existing meta-analysis showing the opposite result, so it appears that the burden is now on the shoulders of researchers who make assumptions about the temporal nature of non-conscious processing without actually testing those assumptions. Conscious intuitions about unconscious processing have often fooled us in the past, as experiments in subliminal perception, problem solving, and visual/auditory illusions have demonstrated for at least a century (Kihlstrom, Barnhardt, & Tataryn, 1992; Hassin 2013).

Additional points brought up in Schwarzkopf's companion response (Schwarzkopf 2014) deserve consideration (*e.g.,* filtering artifacts, publication bias). We refer the reader to the discussion section of the original meta-analysis, where we had already addressed several of these concerns (Mossbridge et al. 2012).

**Acknowledgements**

The authors would like to gratefully acknowledge important suggestions and comments made by Dick Bierman that much improved this paper. In addition, suggests made by an anonymous reviewer at Frontiers in Human Neuroscience also improved the manuscript. Finally, the authors would like to acknowledge the generous funding of the Bial Foundation, which was critical for the production of this manuscript.